\documentclass[]{aa}
\usepackage{graphicx}
\usepackage{txfonts}
\usepackage[authoryear]{natbib}
\bibpunct{(}{)}{;}{a}{}{,}
\newcommand{\src}  {4U\,0115+63}
\newcommand{\ha}  {H$\alpha$}
\newcommand{\ew}  {EW(H$\alpha$)}
\newcommand{\ergs}  {erg s$^{-1}$}

\def\simless{\mathbin{\lower 3pt\hbox
     {$\rlap{\raise 5pt\hbox{$\char'074$}}\mathchar"7218$}}}   
\def\simmore{\mathbin{\lower 3pt\hbox
     {$\rlap{\raise 5pt\hbox{$\char'076$}}\mathchar"7218$}}}   

\begin{document}

   \title{Warped disks during type II outbursts in Be/X-ray
   binaries: evidence from optical polarimetry}

   \subtitle{}
  \author{
        P. Reig\inst{1,2}
        \and
        D. Blinov\inst{1,2,3}
           }

\authorrunning{Reig \& Blinov}
\titlerunning{Polarimetric changes during type II outbursts}

   \offprints{pau@physics.uoc.gr}

\institute{
IESL \& Institute of Astrophysics, Foundation for Research and Technology-Hellas, 
71110 Heraklion, Crete, Greece
\and
University of Crete, Physics Department \& Institute of
Theoretical \& Computational Physics, 71003 Heraklion, Crete, Greece
\and
Astronomical Institute, St. Petersburg State University, Universitetsky pr. 28, Petrodvoretz,
198504, Petersburg, Russia\\
}

   \date{Received ; accepted}

\abstract
{Current models that explain giant (type II) X-ray outbursts in Be/X-ray
binaries (BeXB), are based on the idea of highly distorted disks. They are
believed to occur when a  misaligned and warped disk becomes eccentric, 
allowing the neutron star to capture a large amount of material. The BeXB 4U\,0115+63
underwent two major outbursts in 2015 and 2017.}
{Our aim is to investigate whether the structural changes in the disk 
expected during type II outbursts can be detected through optical polarimetry.} 
{We present the first optical polarimetric observations and new optical spectra 
of the BeXB 4U 0115+63 covering the period 2013--2017. We study in detail 
the shape of the H$\alpha$ line profile and the polarization parameters before, 
during, and after the occurrence of a type II X-ray outburst.
}
{We find significant changes in polarization degree and polarization angle
and highly distorted line profiles during the 2017 X-ray outburst. The degree of
polarization decreased by $\sim$ 1\%, while the polarization angle, which  is
supposed to be related with the disk orientation, first increased by
$\sim 10^{\circ}$ in about two months and then decreased by a similar amount 
and on a similar timescale once the X-ray activity ceased.}
{We interpret the polarimetric and spectroscopic variability as evidence for the presence 
of a warped disk.}


   \maketitle

\section{Introduction}

The X-ray source \src\ is a Be/X-ray binary \citep[BeXB,][]{ziolkowski02,paul11,reig11}. These
systems contain a neutron star and a Be star  in a relatively wide ($P_{\rm
orb}\sim$ few tens of days) and eccentric orbit. The most
prominent feature of a Be star is the gaseous equatorial disk around its
equator. The disk is in Keplerian rotation, is geometrically thin, and is in vertical
hydrostatic equilibrium \citep{rivinius13}.  The disk is of paramount importance
to understanding the behavior of BeXBs as it constitutes the main
source of variability and is responsible for the three main observational
properties of Be stars: emission lines, infrared excess, and polarization. The
emission lines and infrared excess are formed by recombination in the disk.
Linear polarization results from Thomson scattering,  when photons
from the Be star scatter with electrons in the Be disk
\citep{poeckert79,wood96,yudin01,halonen13a,haubois14}.  The effect of Thomson
scattering is to reduce the component of the electric vector parallel to the
scattering plane by $\cos^2\chi$, where $\chi$ is the scattering angle, while
the intensity of the perpendicular component remains unaltered after scattering.
Thus the light becomes polarized perpendicularly to the scattering plane, which
roughly coincides with the plane of the disk.  The polarization angle then gives
information about the orientation of the disk \citep{wood96, quirrenbach97}. The
polarization degree increases with the optical depth (or density) of the gas in
the disk and with the inclination  with respect to the observer
\citep{wood96,halonen13a}.

BeXBs are transient X-ray sources that spend most of the time in a dormant
state, although persistent X-ray sources also exist \citep{reig99}. When active,
transient BeXBs exhibit two types of X-ray outburst. Normal or type I outbursts
show a moderate increase in X-ray flux ($L_X \simless 10^{37}$ \ergs), occur
near periastron passage, and last for a fraction of the orbit. Giant or type II
outbursts are significantly brighter ($L_X \simmore 10^{37}$ \ergs), do not
occur at any preferential orbital phase and may last for several orbits. 

Current models that explain type II outbursts are based on the idea that a
highly misaligned disk becomes warped and eccentric, allowing the neutron star
to capture a large amount of material \citep{martin11,okazaki13,martin14a}.

\citet{okazaki13} performed numerical simulations and showed that the warped
shape of a misaligned disk leads to enhanced mass accretion if the warped part
gets across the orbit of the neutron star. They also showed that  the accretion
rate is much higher for small tilt angles ($\beta \simless 20^{\circ}$) than
for large tilt angles ($\beta \simmore 40^{\circ}$). The reason is that at low
tilt angles, the tidal torque of the neutron star favors the creation of  dense
(denser than the disk) mass streams even before the actual collision between the
disk and the neutron star. When the neutron star interacts with the stream, the
mass accretion rate is enhanced. In highly tilted systems, the stream forms more
slowly because the tidal torque is weaker. Therefore the disk itself has to be
several times denser than in  less tilted systems to supply the same amount of
gas.

\citet{martin14a} showed that giant outbursts can also occur in highly tilted
systems if the disk is highly eccentric. By increasing the eccentricity of the
disk, the neutron star can capture matter from its outer parts more easily than
in a circular disk. There are several ways to make an initially circular disk
eccentric but the most promising one appears to be the Kozai-Lidov mechanism
\citep{kozai62,lidov62,martin14b,fu15a}. The key idea in this mechanism is that
the product of the disk inclination and eccentricity remains constant. Thus, a
test particle that is initially on a circular orbit in a misaligned disk
undergoes a series of oscillations where the inclination and eccentricity
interchange periodically. During the oscillation, the disk can attain large
eccentricities after a few orbital periods. Simulations show that the more
tilted the disk is, the larger the disk can grow and the more eccentric it becomes
\citep{martin14a}.

Since the polarimetric parameters depend so strongly on the internal (density)
and external (orientation) conditions in the disk and the models that explain
type II outbursts require warped, misaligned, eccentric precessing disks, then
we should observe significant changes in the polarization degree and angle
during a giant outburst. This work examines this idea.

\begin{table*}
\caption{Results of the polarization observations of the optical counterpart to 
\src\ in the $R$ band. The uncertainties were determined
following the prescription given in \citet{king14}.}
\label{pol}
\begin{center}
\begin{tabular}{ccccccccc}
\noalign{\smallskip}    \hline\noalign{\smallskip}
JD (2,400,00+) & P[\%] & $\sigma_P$[\%] & PA[deg] & $\sigma_{PA}$[deg] & $q$ &$\sigma_q$ & $u$& $\sigma_u$ \\
\noalign{\smallskip}\hline\noalign{\smallskip}
56549.5671 & 2.8 & 0.7 & -72.6 & 7.4 & -0.0228 & 0.0075 & -0.0158 & 0.0070 \\
56576.4997 & 3.4 & 0.5 & -67.3 & 3.9 & -0.0238 & 0.0047 & -0.0242 & 0.0046 \\
56592.4652 & 3.9 & 0.5 & -67.3 & 3.8 & -0.0272 & 0.0050 & -0.0275 & 0.0052 \\
56623.3618 & 3.0 & 0.3 & -68.8 & 3.1 & -0.0222 & 0.0034 & -0.0202 & 0.0033 \\
56854.5923 & 3.9 & 0.4 & -69.0 & 2.6 & -0.0291 & 0.0035 & -0.0262 & 0.0036 \\
56872.5402 & 4.4 & 0.5 & -66.6 & 3.3 & -0.0301 & 0.0051 & -0.0320 & 0.0050 \\
56885.5121 & 4.2 & 0.5 & -65.3 & 3.5 & -0.0277 & 0.0051 & -0.0321 & 0.0051 \\
56903.5471 & 3.9 & 0.5 & -63.7 & 3.8 & -0.0238 & 0.0053 & -0.0309 & 0.0052 \\
57264.5125 & 3.9 & 0.4 & -70.0 & 2.9 & -0.0301 & 0.0043 & -0.0252 & 0.0037 \\
57346.3532 & 4.2 & 0.2 & -69.0 & 1.8 & -0.0316 & 0.0026 & -0.0284 & 0.0026 \\
57555.5471 & 4.3 & 0.2 & -70.0 & 1.4 & -0.0328 & 0.0022 & -0.0276 & 0.0019 \\
57641.5372 & 3.8 & 0.2 & -68.5 & 1.2 & -0.0279 & 0.0017 & -0.0261 & 0.0016 \\
57665.4385 & 4.0 & 0.1 & -68.1 & 1.0 & -0.0292 & 0.0014 & -0.0279 & 0.0014 \\
57695.4444 & 4.0 & 0.1 & -67.0 & 0.9 & -0.0280 & 0.0013 & -0.0288 & 0.0013 \\
57929.5633 & 4.0 & 0.1 & -70.2 & 0.9 & -0.0312 & 0.0014 & -0.0258 & 0.0012 \\
57957.4767 & 4.0 & 0.1 & -64.9 & 1.0 & -0.0255 & 0.0014 & -0.0306 & 0.0013 \\
57966.4957 & 4.2 & 0.2 & -63.5 & 1.1 & -0.0252 & 0.0017 & -0.0335 & 0.0016 \\
57970.5607 & 4.1 & 0.1 & -66.0 & 1.0 & -0.0277 & 0.0015 & -0.0307 & 0.0015 \\
57985.6152 & 4.2 & 0.2 & -61.5 & 1.4 & -0.0228 & 0.0019 & -0.0352 & 0.0021 \\
57986.5391 & 4.2 & 0.2 & -61.2 & 1.4 & -0.0223 & 0.0020 & -0.0352 & 0.0019 \\
57994.5950 & 3.8 & 0.2 & -62.2 & 1.3 & -0.0215 & 0.0018 & -0.0314 & 0.0018 \\
58015.4837 & 3.9 & 0.3 & -64.3 & 1.9 & -0.0246 & 0.0027 & -0.0309 & 0.0025 \\
58020.5472 & 3.5 & 0.3 & -64.7 & 2.4 & -0.0221 & 0.0029 & -0.0270 & 0.0030 \\
58043.5234 & 3.1 & 0.3 & -70.8 & 3.1 & -0.0246 & 0.0034 & -0.0195 & 0.0034 \\
\noalign{\smallskip}    \hline
\end{tabular}
\end{center}
\end{table*}
\begin{table*}
\caption{Polarization degree in various bands.}
\label{multipol}
\begin{center}
\begin{tabular}{cccccc}
\noalign{\smallskip}    \hline\noalign{\smallskip}
Date &  JD (2,400,000+)     &   $B$  &   $V$   &   $R$  & $I$   \\
        &               &       (\%)    &(\%)   &(\%)   &(\%)  \\
\noalign{\smallskip}    \hline\noalign{\smallskip}
16-06-2016 &57555.56   &4.14$\pm$1.45  &3.89$\pm$0.36  &4.29$\pm$0.21  &3.24$\pm$0.26  \\
10-09-2016 &57641.55   &4.06$\pm$1.43  &4.00$\pm$0.34  &3.82$\pm$0.17  &3.40$\pm$0.15  \\      
02-11-2016 &57695.46   &4.12$\pm$0.59  &3.89$\pm$0.23  &4.02$\pm$0.13  &3.62$\pm$0.15  \\
01-08-2017 &57966.53   &4.12$\pm$0.76  &4.13$\pm$0.25  &4.20$\pm$0.16  &3.88$\pm$0.14   \\
17-10-2017 &58043.54   &4.85$\pm$1.61  &3.07$\pm$0.70  &3.14$\pm$0.34  &2.97$\pm$0.46   \\

\noalign{\smallskip}    \hline
\end{tabular}
\end{center}
\end{table*}

\begin{table*}
\caption{Calibrated photometric magnitudes.  The errors correspond to the standard
deviation of the difference between the measured and the cataloged value of the standard
stars.}
\label{phot}
\begin{center}
\begin{tabular}{cccccc}
\noalign{\smallskip}    \hline \noalign{\smallskip}
Date    &JD (2,400,000+)  & B   &V      &R      &I      \\
\noalign{\smallskip}    \hline \noalign{\smallskip}
29-07-2013 &56503.54 & 16.95$\pm$0.02 & 15.41$\pm$0.02 & 14.48$\pm$0.02 & 13.45$\pm$0.02 \\
29-08-2013 &56534.56 & 17.07$\pm$0.02 & 15.53$\pm$0.02 & 14.62$\pm$0.02 & 13.58$\pm$0.04 \\
20-08-2014 &56890.57 & 16.97$\pm$0.02 & 15.49$\pm$0.02 & 14.61$\pm$0.02 & 13.64$\pm$0.03 \\
14-09-2014 &56915.49 & 16.97$\pm$0.02 & 15.48$\pm$0.01 & 14.58$\pm$0.02 & 13.56$\pm$0.03 \\
22-07-2015 &57226.56 & 16.99$\pm$0.02 & 15.37$\pm$0.01 & 14.37$\pm$0.01 & 13.24$\pm$0.01 \\
18-11-2015 &57345.41 & 16.88$\pm$0.02 & 15.24$\pm$0.01 & 14.22$\pm$0.02 & 13.11$\pm$0.02 \\
08-09-2016 &57640.54 & 16.73$\pm$0.09 & 15.00$\pm$0.08 & 13.91$\pm$0.09 & 12.78$\pm$0.09 \\
06-10-2016 &57668.51 & 16.55$\pm$0.01 & 14.84$\pm$0.01 & 13.76$\pm$0.01 & 12.60$\pm$0.01 \\
03-11-2016 &57696.49 & 16.61$\pm$0.02 & 14.88$\pm$0.02 & 13.79$\pm$0.02 & 12.62$\pm$0.02 \\
25-06-2017 &57930.58 & 16.35$\pm$0.03 & 14.59$\pm$0.03 & 13.47$\pm$0.06 & 12.32$\pm$0.07 \\
11-07-2017 &57946.54 & 16.37$\pm$0.02 & 14.63$\pm$0.02 & 13.57$\pm$0.02 & 12.44$\pm$0.03 \\
28-08-2017 &57994.47 & 16.77$\pm$0.02 & 15.15$\pm$0.02 & 14.12$\pm$0.02 & 13.02$\pm$0.03 \\
\noalign{\smallskip}    \hline
\end{tabular}
\end{center}
\end{table*}

\section{Observations}

All observations reported in this work were obtained using the 1.3 m telescope
at the Skinakas observatory (Crete, Greece). We regularly monitor the optical
continuum ($B$, $V$, $R$, $I$ Jonhson-Cousins bands) and the \ha\ line of \src\
with this telescope since 1999 and the polarization parameters in the $R$ band
since 2013. Here we focus on the observations at the time of the October 2015
and July 2017 X-ray outbursts.

\begin{figure}
\resizebox{\hsize}{!}{\includegraphics{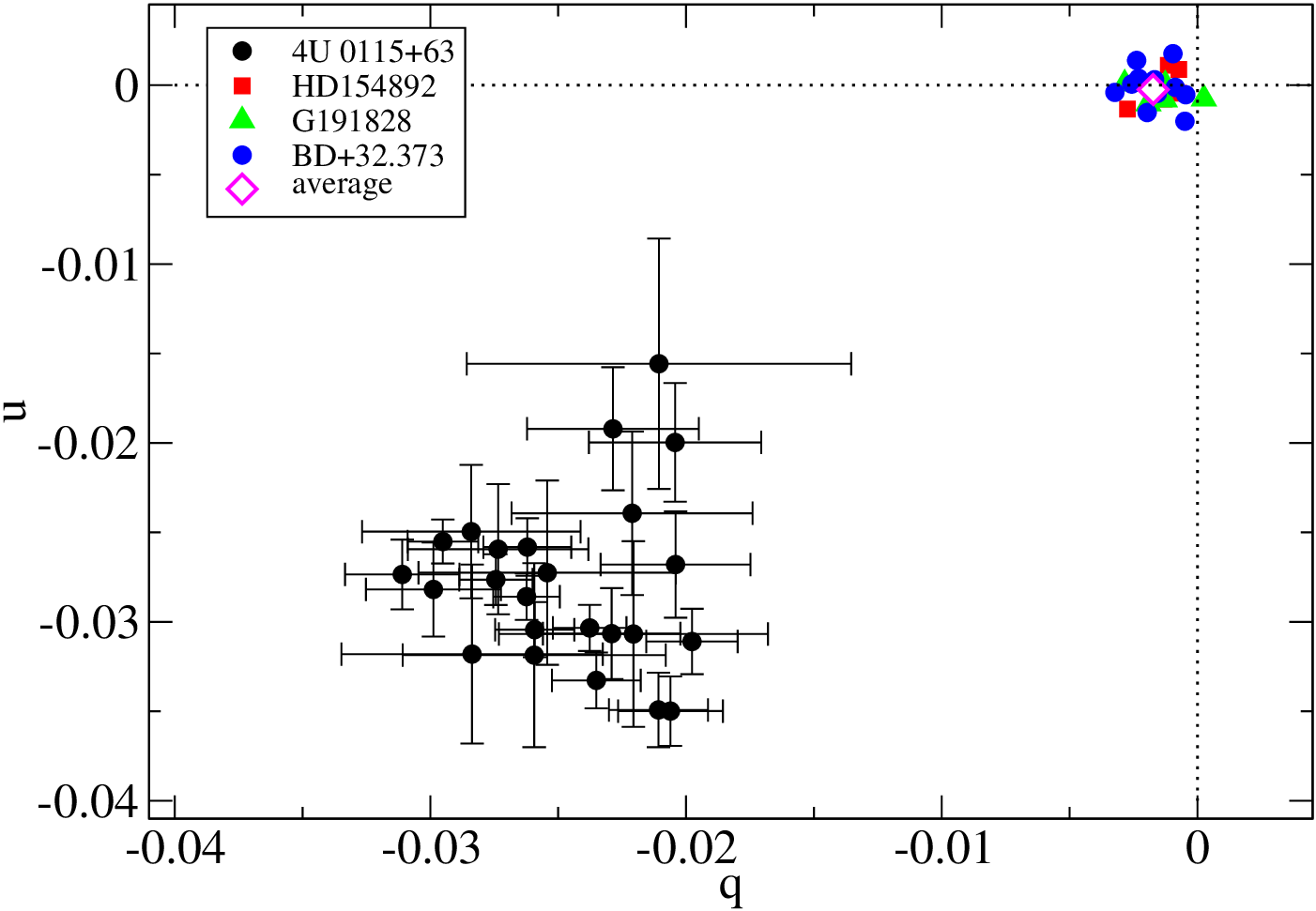}} 
\caption[]{Comparison of the $u$ and $q$ Stokes parameters of \src\ and three 
zero-polarized standard stars in the $R$ band. This figure demonstrates the stability of the
RoboPol polarimeter and rules out any instrumental origin for the source
polarization. The error bars of the standard stars were removed for clarity. The
error bar of the weighted average is smaller than the size of the point. }
\label{instpol}
\end{figure}

\subsection{Polarimetry}

Polarimetric observations were made with the RoboPol polarimeter \citep{king14}.
In the polarimetry configuration, we used a 2048$\times$2048 ANDOR CCD with a
13.5 $\mu$m pixel size. This configuration gives a plate scale of  0.43
$\arcsec$/pixel. RoboPol is an imaging photopolarimeter that uses a combination
of half-wave plates and Wollaston prisms with differing fast axis and prism
orientations. The net result is that every point in the sky is  projected to
four points on the CCD.  The photon counts
in each spot, measured using aperture photometry, are used to calculate the $U$ and $Q$ parameters of linear
polarization. The absence of moving parts allows RoboPol to compute these
parameters in one shot. RoboPol was designed to measure the polarization
parameters of point sources.  The optimization of the instrument sensitivity for
a point source is achieved by using a mask in the telescope focal plane. Because
the mask prevents unwanted photons from the nearby sky and sources from
overlapping with the central target spots, the sky background level surrounding
the central target spots is reduced by a factor of four compared to the field
sources.  The uncertainties of the Stokes parameters and the polarization
degree and angle were determined following the prescription given in
\citet{king14}.  We also observe a number of zero-polarized standards during
each run that were used to derive the instrumental polarization, which
amounts to $B_{\rm inst}=0.38\pm0.02$\%, $V_{\rm inst}=0.35\pm0.09$\%, and 
$I_{\rm inst}=0.86\pm0.14$\%. 

Because the $R$ band is the most frequently used, the correction for the
instrumental polarization in this band is performed in two stages. The details
of the methodology can be found in \citet{king14}; we briefly describe it here: 
the first stage involves measuring the zero-polarized standards in hundreds
of different locations  within the instrument field of view (FoV) several times
per year,  and then fitting their Stokes parameters to find the variation along the
CCD chip. This model allows to correct for the global pattern of instrumental
polarization within the 13' FoV.  However, in each position on the CCD, the
model correction can be systematically off its true value by $0.1-0.2$\%.
Therefore, in order to correct for this systematic and small shift for sources
observed in the mask, we additionally measure zero-polarized standard stars in
the mask.

Figure~\ref{instpol} compares the polarization parameters of the source and
three zero-polarized standard stars. The remaining instrumental polarization
after the model correction is very small compared to the \src\ polarization. The
magenta empty diamond point in Fig.~\ref{instpol} represents the weighted
average ($q=-0.0017\pm0.0002$, $u=-0.0003\pm0.0002$) for the three standards.
Overall, we estimate the amount of instrumental polarization in the $R$ band  to
be $R_{\rm inst}= 0.30\pm0.06$\%.

The results of the polarimetric observations, corrected for instrumental
polarization, are given in Table~\ref{pol} for the $R$ band.  In addition, we
also observed \src\ through $B$, $V$, and $I$ bands on several occasions 
(Table~\ref{multipol}).

\begin{table}
\caption{\ha\ equivalent width and $1-\sigma$ errors of \src. }
\label{spec}
\begin{center}
\begin{tabular}{lcllcccc}
\noalign{\smallskip}    \hline \noalign{\smallskip}
Date    &JD (2,400,000+)  &\ew          \\
        &                 &($\AA$)       \\
\noalign{\smallskip}\hline\noalign{\smallskip}
07-07-2015  &57211.48   &$ -6.9\pm0.3$      \\
15-09-2015  &57281.46   &$ -8.2\pm0.3$      \\
06-09-2016  &57638.51   &$ -9.8\pm0.2$      \\
04-10-2016  &57666.49   &$ -9.6\pm0.3$      \\
12-07-2017  &57947.50   &$-11.6\pm0.4$      \\
13-07-2017  &57948.53   &$-11.9\pm0.5$      \\
29-08-2017  &57995.42   &$-10.3\pm0.6$      \\
07-09-2017  &58004.44   &$ -7.2\pm1.4$      \\
13-10-2017  &58040.32   &$ -4.7\pm0.2$      \\
22-11-2017  &58080.20   &$ -5.8\pm0.4$      \\
\noalign{\smallskip}    \hline
\end{tabular}
\end{center}
\end{table}

\begin{figure}
\resizebox{\hsize}{!}{\includegraphics{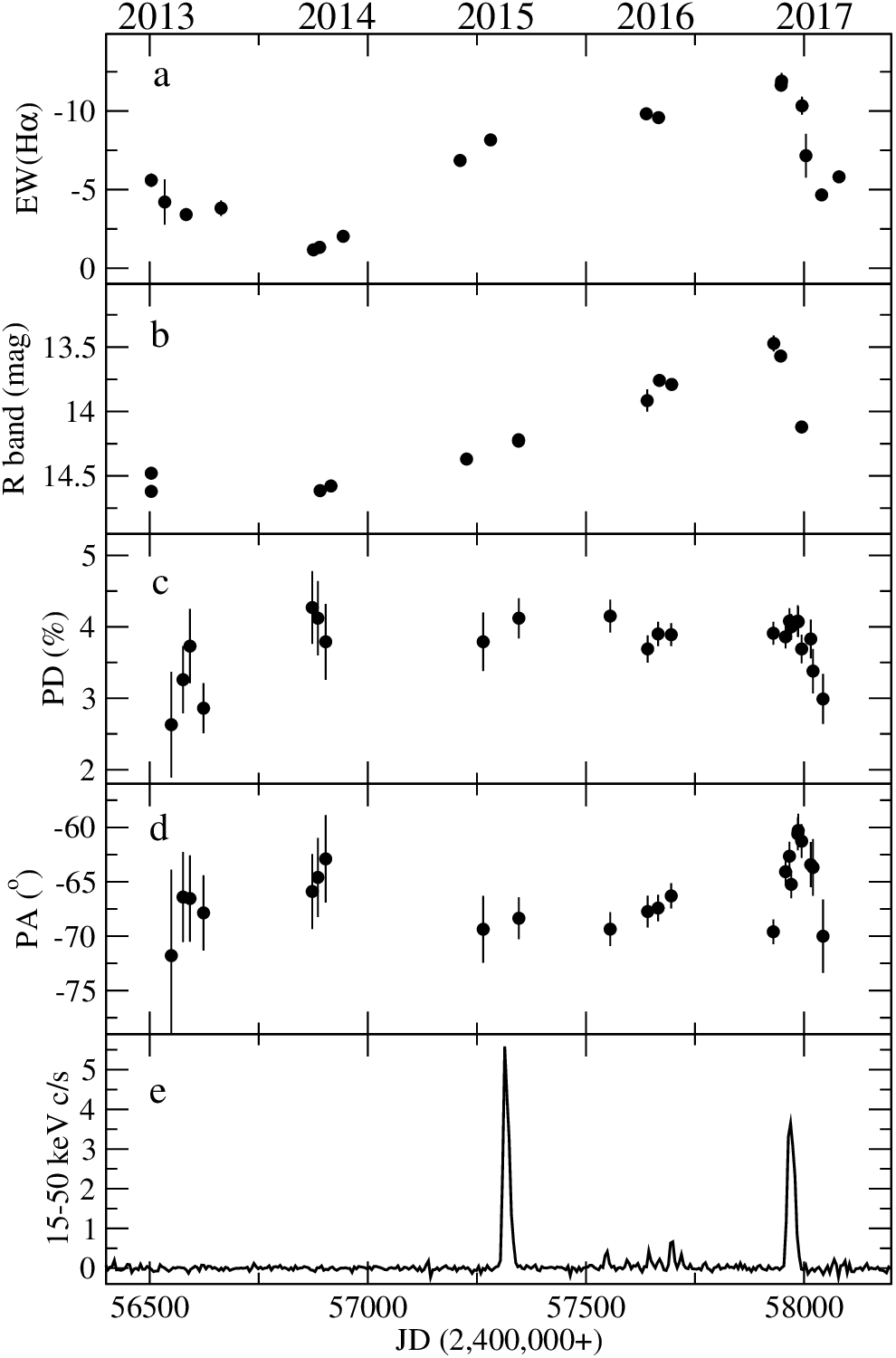}} 
\caption[]{Evolution of the \ha\ equivalent width (a), $R$-band magnitude (b), polarization degree 
in the R-band (c), polarization angle (d), and BAT/{\it Swift} X-ray flux with time.}
\label{obs}
\end{figure}

\subsection{Spectroscopy}

Spectroscopic observations were made with a 2000$\times$800 15 $\mu$m pixel ISA
SITe CCD and a 1302 l~mm$^{-1}$ grating, giving a nominal dispersion of $\sim$1
\AA/pixel until 2016. In 2017, the CCD was replaced by a  2048$\times$2048 13.5
$\mu$m pixel ANDOR IKON with a dispersion of 0.9 \AA/pixel.  Spectra of
comparison lamps were taken before and after each exposure to account for small
variations of the wavelength calibration during the exposure. To ensure
homogeneous processing of the spectra, they were normalized with respect to the
local continuum, which  was rectified to unity by employing a spline fit.  The
error in the \ha\ equivalent width represents the standard deviation of ten
measurements. Each measurement corresponds to a different definition of the
continuum. The results of the spectroscopic observations are summarized in
Table~\ref{spec}.

\subsection{Photometry}

For the photometric observations the telescope was equipped with a 
2048$\times$2048 ANDOR CCD with a 13.5 $\mu$m pixel size, giving a scale of 0.28
$\arcsec$/pixel. Standard stars from the Landolt list \citep{landolt09} were
used for the transformation equations.  Reduction of the data was carried out in
the standard way using the IRAF tools for aperture photometry.    We
calculated the error of the photometry for each night as the standard deviation
of the difference between the observed calibrated magnitudes of the
standard stars and the magnitudes of the catalogue. The results of the
photometric observations are given in Table~\ref{phot}.

\begin{figure}
\resizebox{\hsize}{!}{\includegraphics{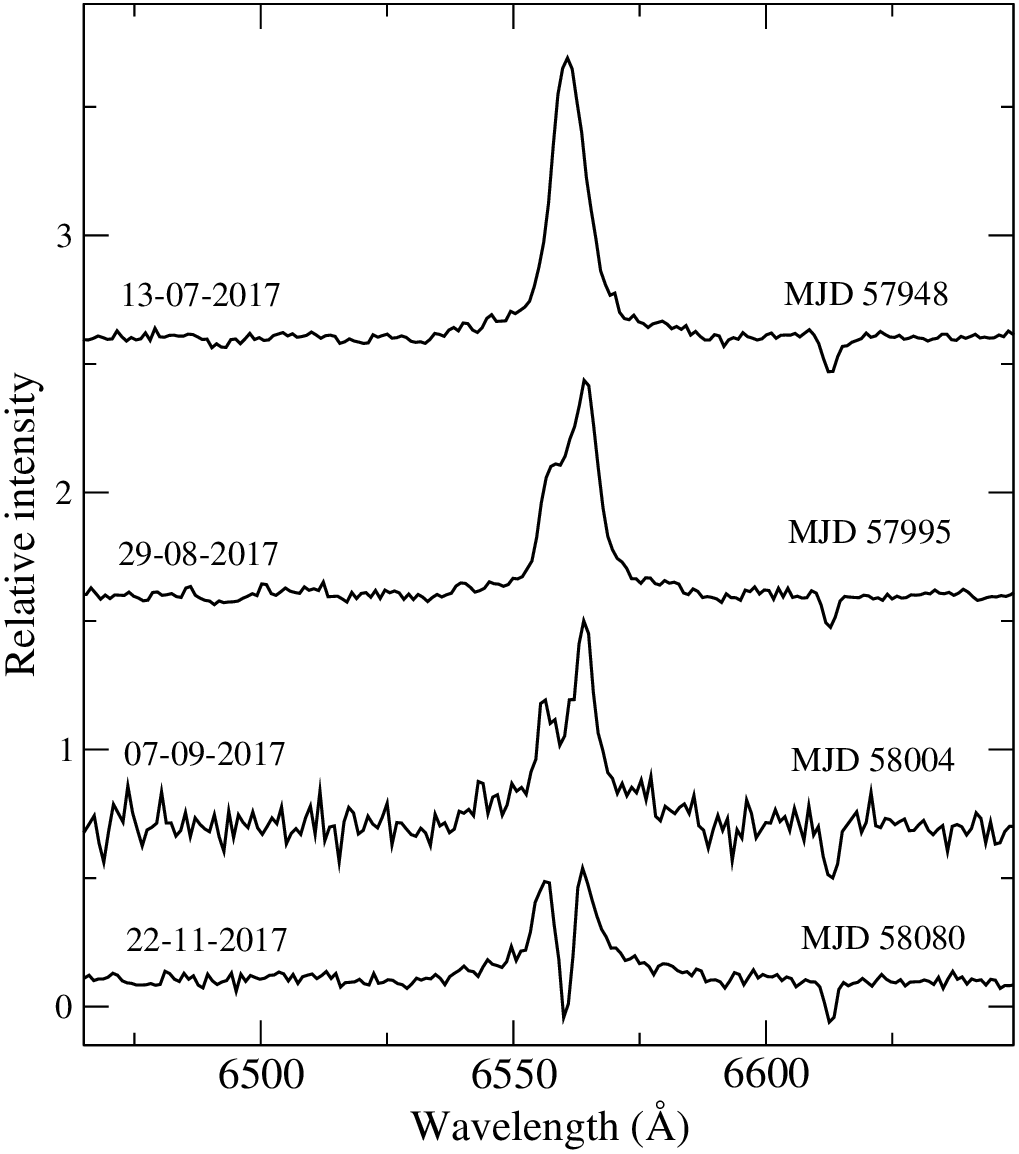}} 
\caption[]{\ha\ line profiles during the July 2017 outburst. }
\label{haprof}
\end{figure}
\begin{figure}
\begin{tabular}{c} 
{\includegraphics[width=8cm,height=5cm]{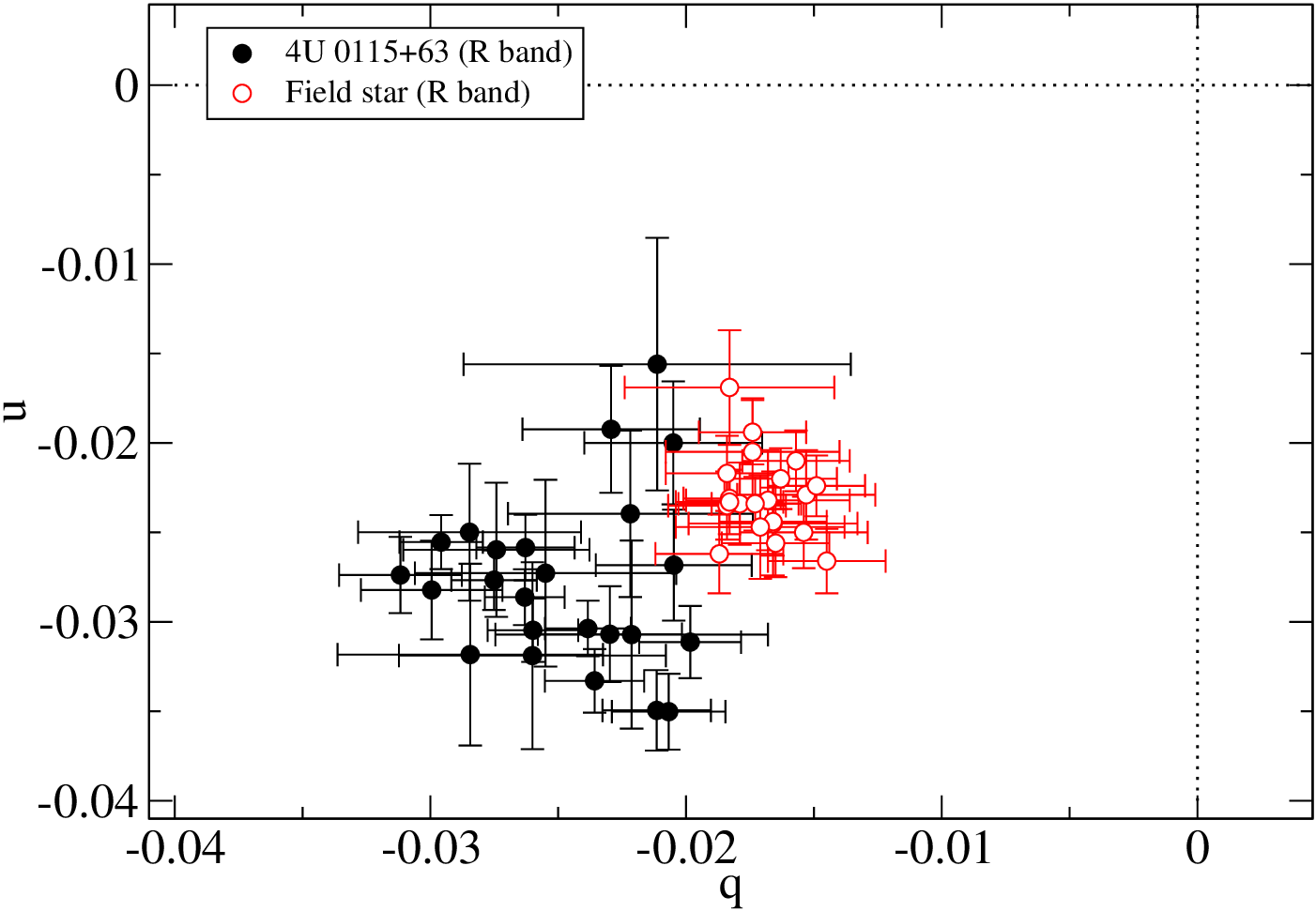}} \\
{\includegraphics[width=8cm,height=5cm]{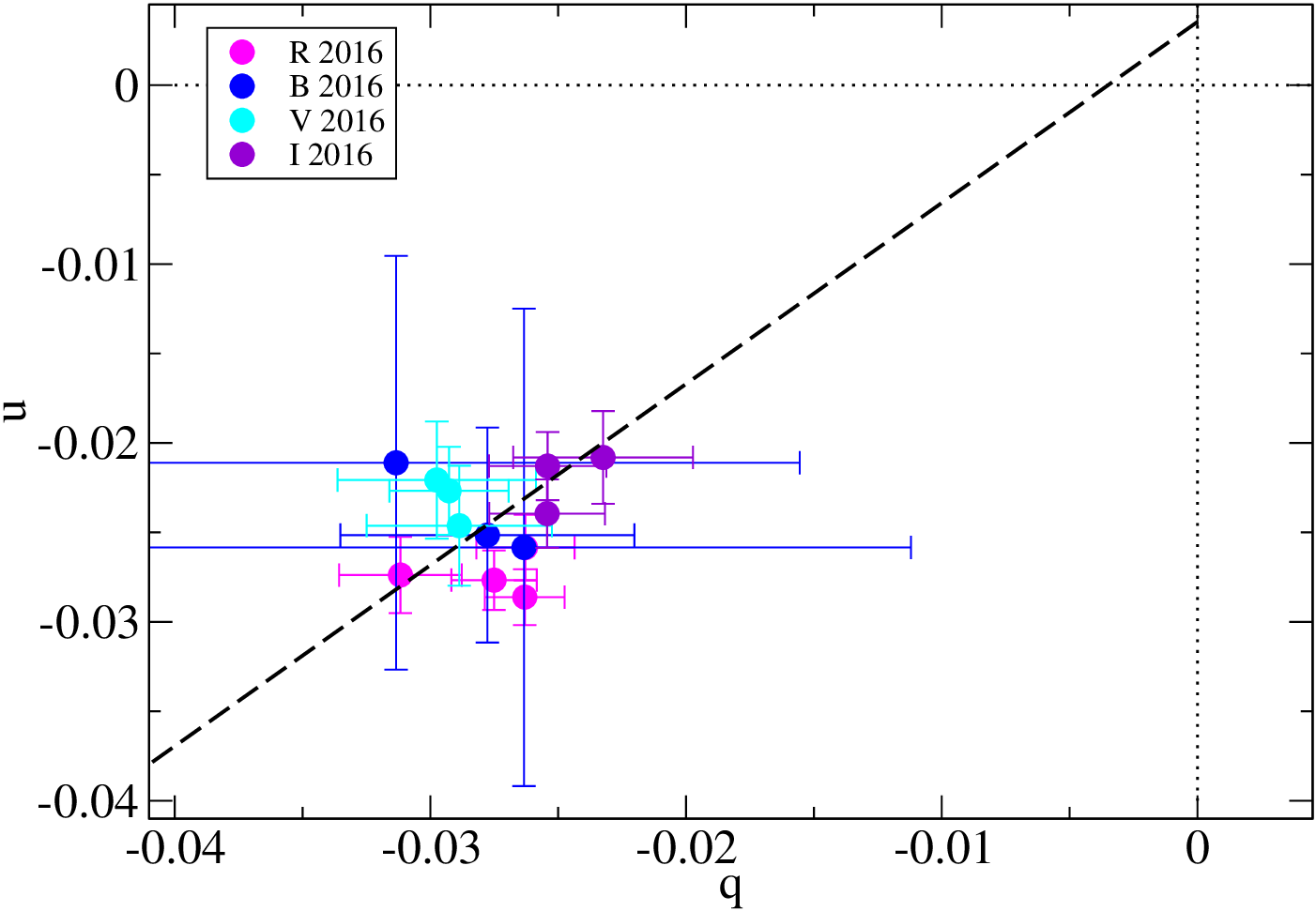}} \\
{\includegraphics[width=8cm,height=5cm]{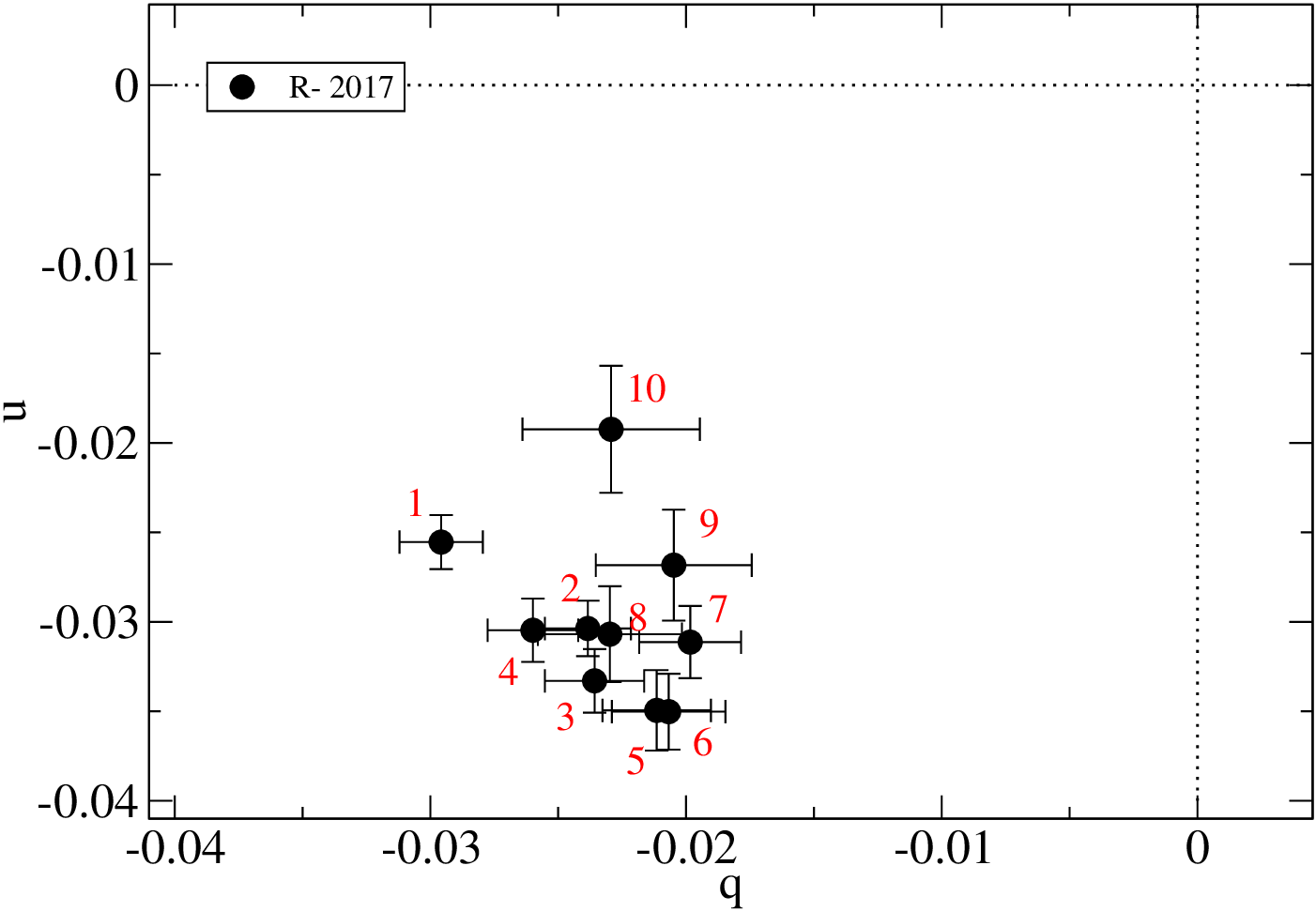}} \\
\end{tabular}
\caption[]{Normalized Stokes parameter diagram. {\it Top panel}: the source
(black filled circles) and a field star (empty red circles)
for observations spanning the period 2013-2017. {\it Middle panel}: source
observations during 2016 (no X-ray outburst). {\it Bottom panel}: 
source observations during the 2017 outburst. The numbers indicate the flow of time, 
where number 1 corresponds to the older observation. }
\label{qu}
\end{figure}

\section{Results}

Figure~\ref{obs} shows the evolution of the \ha\ line equivalent width (\ew),
the $R$-band magnitude, the polarization degree and angle (in the $R$ band), and
the X-ray intensity in the interval 2013-2017. During the period 2014-2016 the
source optical emission experienced a smooth increase both in the continuum and
line components. The source brightened by $\sim$ 1 mag and the \ew\ increased
from 0 to $-11$ \AA. The degree of polarization initially increased by
$\sim$1\%, but it remained rather stable for most of the time.  The optical
parameters did not seem to be affected by the 2015 major X-ray outburst.
However, large changes in all quantities occurred after the 2017 outburst. After
the 2017 outburst, the polarization degree returned to the initial value of
$\sim$3\% (not corrected for interstellar polarization), the polarization angle
returned to pre-outburst values after first increasing and then decreasing by
$\sim10^{\circ}$, and the \ew\ and the optical continuum suffered a fast drop.

The shape of the \ha\ line also experienced large changes as shown in
Fig.~\ref{haprof}.  The 2017 X-ray outburst began on $\sim$ MJD 57960. The
spectrum taken two weeks before, on 13 July 2017 (MJD 57948), exhibits a
single-peak profile and the \ew\ is the largest measured in the past five years.
The maximum X-ray flux was observed on MJD 57970. The spectrum on 29 August
2017  (MJD 57995) shows an asymmetric profile. By 7 September 2017 the asymmetry
had grown and a red-dominated double-peak line was apparent. The X-ray flux
returned to pre-outburst values on MJD 57996. The \ha\ line profile of the
spectrum taken on 13 October 2017 (MJD 58040) and 22 November 2017 (MJD 58080)
has the shape of a shell line.

The polarization measurements presented in Tables~\ref{pol} and \ref{multipol}
and in Fig.~\ref{obs} are corrected for instrumental polarization but  they
are not corrected for the contribution from the interstellar medium (ISM).
Estimating this contribution is not easy. \src\ is located in the Galactic plane
(galactic latitude $b=1^{\circ}$) at a large distance, $\sim 6-7$ kpc
\citep{negueruela01,reig15}. Hence the number of molecular clouds that lie
between us and the source and how each one of them affects the polarization
properties is unknown. The good news is that the polarization introduced by the
ISM does not affect the results and conclusions of this work.  The reason is
that the interstellar component of the polarization for a given direction in the
sky is constant with time at the time scales considered in this work. Therefore the
variability observed in 2017  indicates that the polarization changes are
intrinsic to the source. 

Further evidence that some fraction of the measured polarization is
intrinsic to the source can be obtained from the normalized Stokes parameter
plane, the so-called $q-u$ diagram, shown in Fig.~\ref{qu}. The top panel of
this figure shows the Stokes parameters of the source and a field star. The
smaller scatter of the field star observations supports the idea that the \src\
is variable. The middle and bottom panels of  Fig.~\ref{qu} show the $q$ and $u$
measurements of \src\ at two different epochs: during an X-ray quiet state (2016
observations) and during to the 2017 giant outburst.   Because a star that is
variable in degree but constant in position angle would move along a line in
this plot, any elongation of the distribution of data points along a general
direction in this diagram is good evidence for intrinsic polarization
\citep{clarke87,mcdavid99}. The dashed line in Fig.~\ref{qu} (middle panel) is
the best fit to the data points. The angle between this line and the $q$ axis
defines a preferred direction associated with the star, which gives a good
approximation to the polarization angle of the intrinsic polarization. The fact
that the line does not pass through the origin ($q=0$, $u=0$) indicates that
there is an intrinsic component. On the other hand, the fact that the distance
from the origin to the line is very short indicates that it is difficult to
disentangle the interstellar component from the intrinsic one because their
position angles are nearly the same. In contrast,  the   $q-u$ diagram of the
2017 observations  deviates from a straight line; instead, it makes a loop in
which the source moves counter-clockwise. This shape clearly indicates that not
only the degree of polarization but also the polarization angle changed.

\begin{figure}
\resizebox{\hsize}{!}{\includegraphics{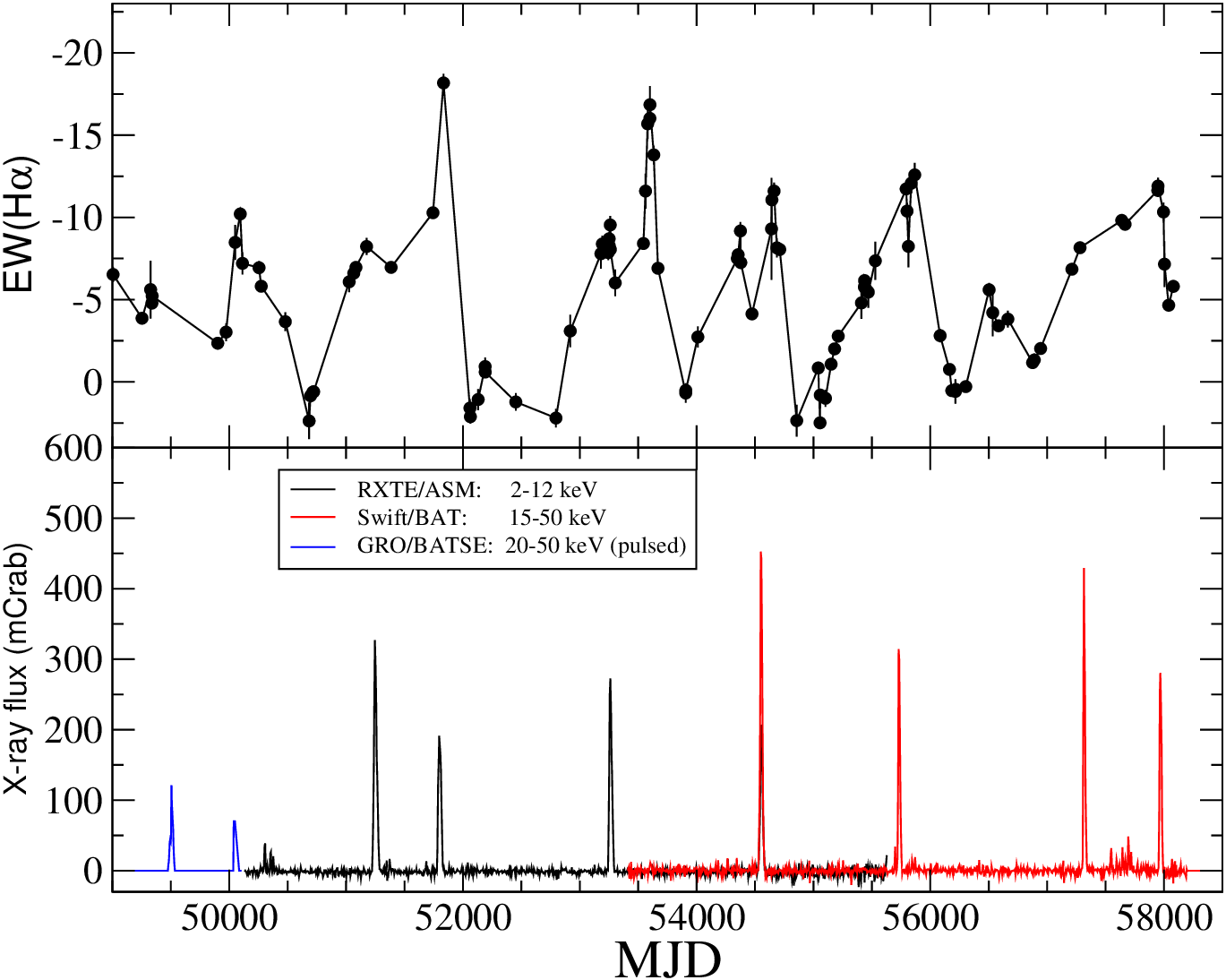}} 
\caption[]{Long-term evolution of the \ha\ equivalent width of \src\ in relation
to its X-ray activity. Each spike in the bottom panel corresponds to a giant
outburst. Optical data prior to 2015 come from \citet{reig16}. From
2015 onwards the data points represent new observations obtained with the 1.3 m
telescope at the Skinakas observatory. }
\label{mcrab}
\end{figure}

\section{Discussion}

In this section we discuss the implications of the observations, with particular
emphasis on the observations around the 2017 outburst and provide an
interpretation of the dramatic changes that occurred during that event.

The data presented in this work show that the source follows the same pattern of
variability described in \citet{negueruela01} and  \citet{reig07}, namely the
quasi-periodic nature of the giant X-ray outbursts and the cyclic line profile
variability observed in \src. The recurrence time of type II outbursts of about
3 years  was already pointed out by \citet{whitlock89} and confirmed by
subsequent studies \citep{negueruela01}. \citet{reig07} realised that sometimes
the three-year cycle is broken by the presence of two more closely separated (1--1.5
years) outbursts. When this occurs, the \ew\ prior to the start of the second
outburst is always larger than that of the first. Likewise, the peak
X-ray flux of the second outburst of the pair is lower. The optical parameters do
not significantly change after the first outburst. In contrast, after the second
outburst, the disk suffers large structural changes that lead to its complete
dissipation. This is illustrated in Fig.~\ref{mcrab}\footnote{We note that given
the different energy ranges and characteristics of the instrument a comparison
of the intensity of the outbursts is only meaningful for the same instrument.},
which shows long-term correlated optical and X-ray behavior of the system.

\begin{figure}
\includegraphics[width=7cm,height=8.5cm]{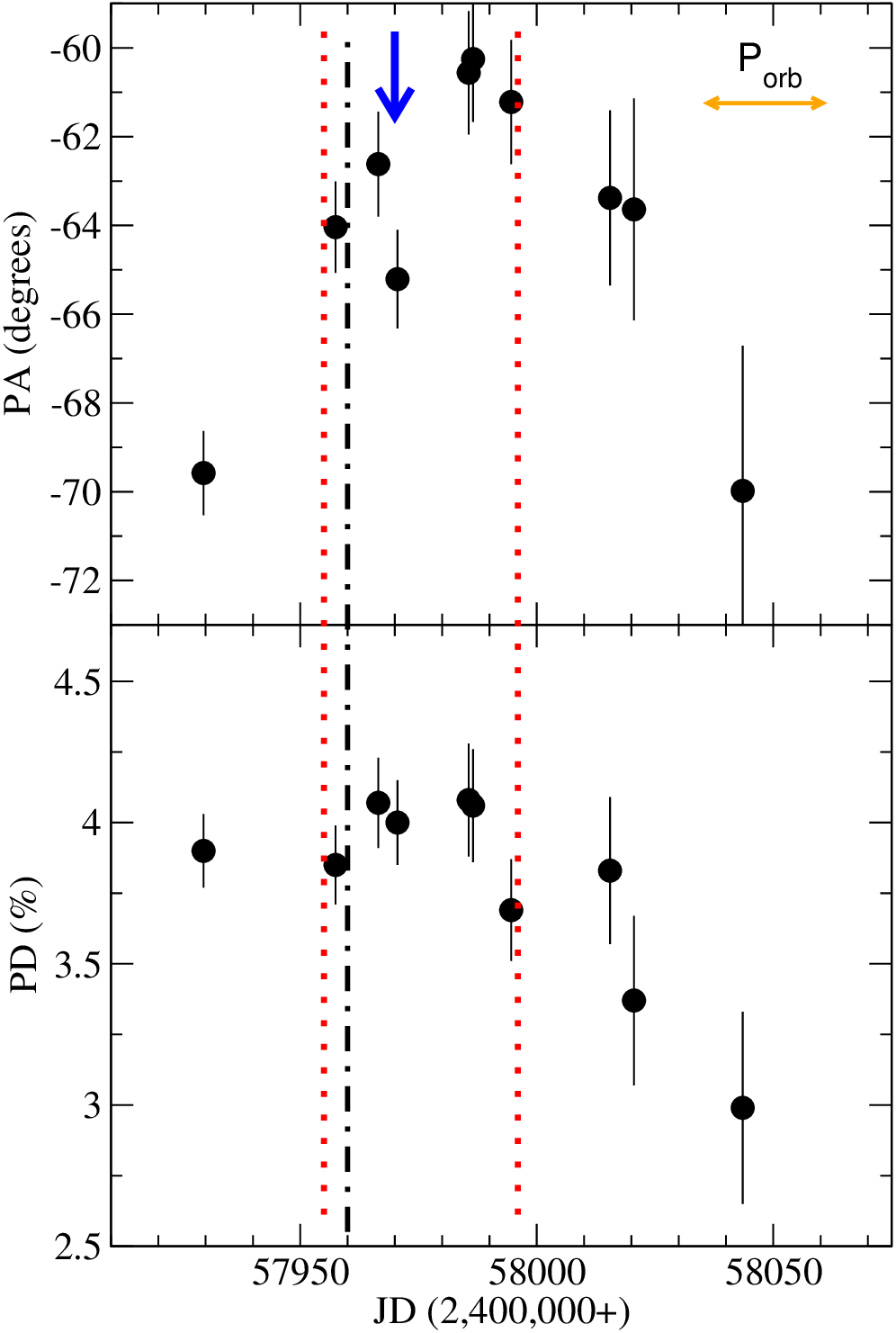} 
\caption[]{Polarization angle and degree of polarization during the 2017 giant
X-ray outburst. The dotted red lines mark the beginning
and end of the outburst,  the blue arrow indicates the X-ray maximum
intensity, and the dashed-dotted black line shows the time at periastron. }
\label{outburst}
\end{figure}

The period spanned by our observations includes two outbursts, one in October 2015
and another in August 2017. Unfortunately, the coverage of the 2015 outburst
was very poor. With only two data points it is difficult to draw any firm
conclusions. However, although there might be an observational bias, the absence
of large changes during this outburst agrees very well with what is observed
in previous events. As can be seen in Fig.~\ref{mcrab}, the \ew\ remained fairly
unaltered after the 1994 (MJD 49500) and the 1998 (MJD 51260) outbursts. Similarly neither
the line (\ew) nor the continuum emission ($R$ mag) were greatly affected after the
2015 (MJD 57320) outburst (Fig.~\ref{obs}). In contrast, Figs. ~\ref{haprof}
and ~\ref{outburst} show the dramatic changes that occurred in the Be disk at
the time of the 2017 giant outburst. 

Figure~\ref{haprof} displays some characteristic profiles of the \ha\
immediately before, during, and after the 2017 outburst.  The line profile
changed from a strong single-peak symmetric profile immediately before the onset
of the outburst to a highly asymmetric profile during the outburst and a
shell-like profile once the X-ray activity had ceased. Be stars exhibit a wealth
of line profiles. Generally they fall in two categories, symmetric and
asymmetric profiles \citep{hummel94,hummel95,hanuschik96a,hummel97,silaj10}. 
Symmetric profiles are generated in stable Keplerian disk configurations.
Asymmetric profiles are believed to arise from distorted disks. Symmetric
profiles include single-peak, double-peak, wine-bottle (single or double peak
profiles that show inflections of the shoulders of the line), and  shell lines
(double-peak line with the central depression below the stellar continuum).
Double-peak lines arise at intermediate inclination angles and result from
Doppler shifts of the particles moving in the disk. Single-peak and wine-bottle
lines occur when the observer sees the disk at low inclination angle or when
the disk is large enough (the peak separation decreases as the disk radius
increases) that the spectral resolution cannot separate the line into two peaks.
Shell lines are thought to occur in edge-on systems with the central intensity
resulting from partial absorption of the central star by the disk
\citep{hanuschik95,rivinius06,silaj14}. Asymmetric lines are characterized by
double-peak profiles in which one of the peaks is stronger than the other. The
relative intensity of the two peaks may change over time giving rise to the
so-called V/R variability, which is believed to be caused by density
perturbations in the disk \citep{telting94}.

Because the spin axis of the Be star cannot change on timescales of days, the
appearance of different profiles in the  same star can only imply that the
disk axis is changing direction. Indeed the changes in the emission line profile
in Be and BeXBs have been interpreted as observational evidence for misaligned
disks that become warped in their outer parts 
\citep{hummel98,negueruela01,reig07,moritani11,moritani13}.

Figure~\ref{outburst} shows a detailed view of the polarization degree and
polarization angle during the 2017 outburst. The dotted red lines mark the
beginning and end of the outburst, while the dashed-dotted black line corresponds
to periastron \citep{raichur10}. The peak of the outburst as seen by {\it
Swift}/BAT is shown by the blue arrow. The polarization degree decreased by
$\sim$1\% after the outburst. The polarization angle first increased by
$\sim10^{\circ}$ and then decreased by the same amount. The larger changes
occurred before and after the X-ray outburst. In Thomson scattering, the
polarization angle of the resulting polarized light is perpendicular to the
scattering plane (the plane containing the incident and scattered radiation).
Since the photons that get scattered come from the Be star and the scattering
medium is the disk, the polarization angle is expected to be perpendicular to
the major elongation axis. This result has been confirmed by interferometric
observations \citep{quirrenbach97}. Therefore we conclude that the observed
variation of the polarization angle is related to changes in the orientation of
the disk. 

This is the first time that such a rapid change in polarization parameters is
reported for a BeXB. The position angle measured by polarization has not been
reported to vary in isolated Be stars either, but some exceptions are known, the
most notable being that of the Be star Pleione \citep{hirata07}. The
polarization angle of this star changed from $\sim60^{\circ}$ to
$\sim130^{\circ}$ in the interval 1974--2003. This change was interpreted in
terms of the precession of the disk as anticipated by \citet{hummel98}. 

Regarding BeXB, the only system for which a detailed polarimetric study has been
performed is X-Per \citep{kunjaya95,roche97}. The data covered a declining phase
in the evolution of the disk, and the subsequent recovery. The polarization
degree correlated well with other optical parameters such as the V magnitude and
the \ew. During the disk-loss episode, all three parameters reached a minimum
value. 

The type of variation that we report here for \src\ is rather different from
these two cases. First, the timescales are remarkably faster in \src. The
precession period in Pleione was estimated to be 81 years. The variability
observed in polarization in X-per occurred over several years, in agreement with
the viscous timescales expected in a decretion disk when it goes through a
dissipation/formation phase. In \src,  the polarization angle changed by
$\sim10^{\circ}$ in less than two months. Second, during the declining  phase,
X-Per moved from the lower-right part to the upper-left part in the Stokes
normalized plane, and followed the same trace but in opposite direction during
the reformation phase. This means that X-Per followed a straight line
\citep{kunjaya95,roche97}, indicating that the polarization angle did not change
significantly. As shown in Fig.~\ref{qu}, the pattern traced by \src\ in the
normalized Stokes parameter plane is curved, indicating changes both in
polarization degree and angle. Unlike \src, X-Per does not show type II
outbursts, but occasional small-amplitude and slow increases in X-rays. Hence we
conclude that type II outbursts appear as violent events that lead to dramatic
changes in the disk.

Figure~\ref{outburst} conveys another very interesting result. The first data
point shown in this figure was  taken approximately one orbital period ($P_{\rm
orb}=24.3$ days) before the onset of the outburst (left red dotted line), which
in turn,  roughly coincides with the time at which the neutron star crossed
periastron  (dashed-dotted black line), according to the ephemeris of
\citet{raichur10}. Although we cannot be sure of when the disk began to warp due
to the lack of observations prior to the X-ray outburst, the fact that the
polarization angle of the first 2017 observation is consistent with the
long-term average ($-67\pm2^{\circ}$) suggests that the disk then still
presented a stable configuration. However, soon after (one orbital period before
the start of the outburst), the tidal torque exerted by the neutron star
perturbed the disk, which began to warp. This event changed the orientation of
the disk and gave rise to the observed changes in the polarization angle. In the
next passage, the neutron star moved across the warped (outer) part of
the disk resulting in enhanced mass accretion and the X-ray emission.  The disk
axis began to precess, completing an entire cycle in $\sim120$ days. This is a
relatively very short timescale, but is consistent with  the relatively small
variation in the polarization angle. Based on spectroscopic observations,
\citet{moritani13} estimated the precession of a warped disk in the BeXB
1A\,0535+262 to be several hundred days. A detailed spectroscopic and
polarimetric monitoring of this source and other BeXBs during future giant
outbursts will clarify whether our interpretation is correct and will allow  for
more accurate estimations of the timescales associated with disk warping in
BeXBs.

\section{Conclusion}

We report for the first time changes in the optical polarization parameters
during a giant X-ray outburst in the BeXB \src. Most notably, the polarization
angle, and the profile of the \ha\ line, and therefore the orientation of the Be disk,
changed on timescales comparable to the orbital period. We interpret this
variability as evidence for a warped disk, supporting models that predict highly
perturbed disks as the origin of type II outbursts in BeXB.

\begin{acknowledgements}

Skinakas Observatory is run by the University of Crete and the Foundation for
Research and Technology-Hellas. 

\end{acknowledgements}

\bibliographystyle{aa}
\bibliography{../../../pol}

\begin{thebibliography}{43}
\expandafter\ifx\csname natexlab\endcsname\relax\def\natexlab#1{#1}\fi

\bibitem[{{Clarke} \& {McGale}(1987)}]{clarke87}
{Clarke}, D. \& {McGale}, P.~A. 1987, \aap, 178, 294

\bibitem[{{Fu} {et~al.}(2015){Fu}, {Lubow}, \& {Martin}}]{fu15a}
{Fu}, W., {Lubow}, S.~H., \& {Martin}, R.~G. 2015, \apj, 807, 75

\bibitem[{{Halonen} {et~al.}(2013){Halonen}, {Mackay}, \& {Jones}}]{halonen13a}
{Halonen}, R.~J., {Mackay}, F.~E., \& {Jones}, C.~E. 2013, \apjs, 204, 11

\bibitem[{{Hanuschik}(1995)}]{hanuschik95}
{Hanuschik}, R.~W. 1995, \aap, 295, 423

\bibitem[{{Hanuschik}(1996)}]{hanuschik96a}
{Hanuschik}, R.~W. 1996, \aap, 308, 170

\bibitem[{{Haubois} {et~al.}(2014){Haubois}, {Mota}, {Carciofi}, {Draper},
  {Wisniewski}, {Bednarski}, \& {Rivinius}}]{haubois14}
{Haubois}, X., {Mota}, B.~C., {Carciofi}, A.~C., {et~al.} 2014, \apj, 785, 12

\bibitem[{{Hirata}(2007)}]{hirata07}
{Hirata}, R. 2007, in Astronomical Society of the Pacific Conference Series,
  Vol. 361, Active OB-Stars: Laboratories for Stellare and Circumstellar
  Physics, ed. A.~T. {Okazaki}, S.~P. {Owocki}, \& S.~{Stefl}, 267

\bibitem[{{Hummel}(1994)}]{hummel94}
{Hummel}, W. 1994, \aap, 289, 458

\bibitem[{{Hummel}(1998)}]{hummel98}
{Hummel}, W. 1998, \aap, 330, 243

\bibitem[{{Hummel} \& {Hanuschik}(1997)}]{hummel97}
{Hummel}, W. \& {Hanuschik}, R.~W. 1997, \aap, 320, 852

\bibitem[{{Hummel} \& {Vrancken}(1995)}]{hummel95}
{Hummel}, W. \& {Vrancken}, M. 1995, \aap, 302, 751

\bibitem[{{King} {et~al.}(2014){King}, {Blinov}, {Ramaprakash}, {Myserlis},
  {Angelakis}, {Balokovi{\'c}}, {Feiler}, {Fuhrmann}, {Hovatta}, {Khodade},
  {Kougentakis}, {Kylafis}, {Kus}, {Modi}, {Paleologou}, {Panopoulou},
  {Papadakis}, {Papamastorakis}, {Paterakis}, {Pavlidou}, {Pazderska},
  {Pazderski}, {Pearson}, {Rajarshi}, {Readhead}, {Reig}, {Steiakaki},
  {Tassis}, \& {Zensus}}]{king14}
{King}, O.~G., {Blinov}, D., {Ramaprakash}, A.~N., {et~al.} 2014, \mnras, 442,
  1706

\bibitem[{{Kozai}(1962)}]{kozai62}
{Kozai}, Y. 1962, \aj, 67, 591

\bibitem[{{Kunjaya} \& {Hirata}(1995)}]{kunjaya95}
{Kunjaya}, C. \& {Hirata}, R. 1995, \pasj, 47, 589

\bibitem[{{Landolt}(2009)}]{landolt09}
{Landolt}, A.~U. 2009, \aj, 137, 4186

\bibitem[{{Lidov}(1962)}]{lidov62}
{Lidov}, M.~L. 1962, \planss, 9, 719

\bibitem[{{Martin} {et~al.}(2014{\natexlab{a}}){Martin}, {Nixon}, {Armitage},
  {Lubow}, \& {Price}}]{martin14a}
{Martin}, R.~G., {Nixon}, C., {Armitage}, P.~J., {Lubow}, S.~H., \& {Price},
  D.~J. 2014{\natexlab{a}}, \apjl, 790, L34

\bibitem[{{Martin} {et~al.}(2014{\natexlab{b}}){Martin}, {Nixon}, {Lubow},
  {Armitage}, {Price}, {Do{\u g}an}, \& {King}}]{martin14b}
{Martin}, R.~G., {Nixon}, C., {Lubow}, S.~H., {et~al.} 2014{\natexlab{b}},
  \apjl, 792, L33

\bibitem[{{Martin} {et~al.}(2011){Martin}, {Pringle}, {Tout}, \&
  {Lubow}}]{martin11}
{Martin}, R.~G., {Pringle}, J.~E., {Tout}, C.~A., \& {Lubow}, S.~H. 2011,
  \mnras, 416, 2827

\bibitem[{{McDavid}(1999)}]{mcdavid99}
{McDavid}, D. 1999, \pasp, 111, 494

\bibitem[{{Moritani} {et~al.}(2011){Moritani}, {Nogami}, {Okazaki}, {Imada},
  {Kambe}, {Honda}, {Hashimoto}, \& {Ichikawa}}]{moritani11}
{Moritani}, Y., {Nogami}, D., {Okazaki}, A.~T., {et~al.} 2011, \pasj, 63, 25

\bibitem[{{Moritani} {et~al.}(2013){Moritani}, {Nogami}, {Okazaki}, {Imada},
  {Kambe}, {Honda}, {Hashimoto}, {Mizoguchi}, {Kanda}, {Sadakane}, \&
  {Ichikawa}}]{moritani13}
{Moritani}, Y., {Nogami}, D., {Okazaki}, A.~T., {et~al.} 2013, \pasj, 65, 83

\bibitem[{{Negueruela} {et~al.}(2001){Negueruela}, {Okazaki}, {Fabregat},
  {Coe}, {Munari}, \& {Tomov}}]{negueruela01}
{Negueruela}, I., {Okazaki}, A.~T., {Fabregat}, J., {et~al.} 2001, \aap, 369,
  117

\bibitem[{{Okazaki} {et~al.}(2013){Okazaki}, {Hayasaki}, \&
  {Moritani}}]{okazaki13}
{Okazaki}, A.~T., {Hayasaki}, K., \& {Moritani}, Y. 2013, \pasj, 65, 41

\bibitem[{{Paul} \& {Naik}(2011)}]{paul11}
{Paul}, B. \& {Naik}, S. 2011, Bulletin of the Astronomical Society of India,
  39, 429

\bibitem[{{Poeckert} {et~al.}(1979){Poeckert}, {Bastien}, \&
  {Landstreet}}]{poeckert79}
{Poeckert}, R., {Bastien}, P., \& {Landstreet}, J.~D. 1979, \aj, 84, 812

\bibitem[{{Quirrenbach} {et~al.}(1997){Quirrenbach}, {Bjorkman}, {Bjorkman},
  {Hummel}, {Buscher}, {Armstrong}, {Mozurkewich}, {Elias}, \&
  {Babler}}]{quirrenbach97}
{Quirrenbach}, A., {Bjorkman}, K.~S., {Bjorkman}, J.~E., {et~al.} 1997, \apj,
  479, 477

\bibitem[{{Raichur} \& {Paul}(2010)}]{raichur10}
{Raichur}, H. \& {Paul}, B. 2010, \mnras, 406, 2663

\bibitem[{{Reig}(2011)}]{reig11}
{Reig}, P. 2011, \apss, 332, 1

\bibitem[{{Reig} \& {Coe}(1999)}]{reig99}
{Reig}, P. \& {Coe}, M.~J. 1999, \mnras, 302, 700

\bibitem[{{Reig} \& {Fabregat}(2015)}]{reig15}
{Reig}, P. \& {Fabregat}, J. 2015, \aap, 574, A33

\bibitem[{{Reig} {et~al.}(2007){Reig}, {Larionov}, {Negueruela}, {Arkharov}, \&
  {Kudryavtseva}}]{reig07}
{Reig}, P., {Larionov}, V., {Negueruela}, I., {Arkharov}, A.~A., \&
  {Kudryavtseva}, N.~A. 2007, \aap, 462, 1081

\bibitem[{{Reig} {et~al.}(2016){Reig}, {Nersesian}, {Zezas}, {Gkouvelis}, \&
  {Coe}}]{reig16}
{Reig}, P., {Nersesian}, A., {Zezas}, A., {Gkouvelis}, L., \& {Coe}, M.~J.
  2016, \aap, 590, A122

\bibitem[{{Rivinius} {et~al.}(2013){Rivinius}, {Carciofi}, \&
  {Martayan}}]{rivinius13}
{Rivinius}, T., {Carciofi}, A.~C., \& {Martayan}, C. 2013, \aapr, 21, 69

\bibitem[{{Rivinius} {et~al.}(2006){Rivinius}, {{\v S}tefl}, \&
  {Baade}}]{rivinius06}
{Rivinius}, T., {{\v S}tefl}, S., \& {Baade}, D. 2006, \aap, 459, 137

\bibitem[{{Roche} {et~al.}(1997){Roche}, {Larionov}, {Tarasov}, {Fabregat},
  {Clark}, {Coe}, {Kalv}, {Larionova}, {Negueruela}, {Norton}, \&
  {Reig}}]{roche97}
{Roche}, P., {Larionov}, V., {Tarasov}, A.~E., {et~al.} 1997, \aap, 322, 139

\bibitem[{{Silaj} {et~al.}(2014){Silaj}, {Jones}, {Sigut}, \&
  {Tycner}}]{silaj14}
{Silaj}, J., {Jones}, C.~E., {Sigut}, T.~A.~A., \& {Tycner}, C. 2014, \apj,
  795, 82

\bibitem[{{Silaj} {et~al.}(2010){Silaj}, {Jones}, {Tycner}, {Sigut}, \&
  {Smith}}]{silaj10}
{Silaj}, J., {Jones}, C.~E., {Tycner}, C., {Sigut}, T.~A.~A., \& {Smith}, A.~D.
  2010, \apjs, 187, 228

\bibitem[{{Telting} {et~al.}(1994){Telting}, {Heemskerk}, {Henrichs}, \&
  {Savonije}}]{telting94}
{Telting}, J.~H., {Heemskerk}, M.~H.~M., {Henrichs}, H.~F., \& {Savonije},
  G.~J. 1994, \aap, 288, 558

\bibitem[{{Whitlock} {et~al.}(1989){Whitlock}, {Roussel-Dupre}, \&
  {Priedhorsky}}]{whitlock89}
{Whitlock}, L., {Roussel-Dupre}, D., \& {Priedhorsky}, W. 1989, \apj, 338, 381

\bibitem[{{Wood} {et~al.}(1996){Wood}, {Bjorkman}, {Whitney}, \&
  {Code}}]{wood96}
{Wood}, K., {Bjorkman}, J.~E., {Whitney}, B.~A., \& {Code}, A.~D. 1996, \apj,
  461, 828

\bibitem[{{Yudin}(2001)}]{yudin01}
{Yudin}, R.~V. 2001, \aap, 368, 912

\bibitem[{{Ziolkowski}(2002)}]{ziolkowski02}
{Ziolkowski}, J. 2002, \memsai, 73, 1038

\end{thebibliography}

\end{document}